# A closed-loop AI framework for hypothesis-driven and interpretable materials design

Kangyu Ji[1,2‡*], Tianran Liu[1,2‡*], Fang Sheng[1], Shaun Tan[3], Moungi Bawendi[3], Tonio Buonassisi[1*]

[1]Department of Mechanical Engineering, Massachusetts Institute of Technology, 77 Massachusetts Ave, Cambridge, Massachusetts, 02139, United States

[2]Research Laboratory of Electronics, Massachusetts Institute of Technology, 77 Massachusetts Ave, Cambridge, Massachusetts, 02139, United States

[3]Department of Chemistry, Massachusetts Institute of Technology, 77 Massachusetts Avenue, Cambridge, Massachusetts, 02139, United States

[‡]These authors contributed equally to this work.

*Corresponding to Kangyu Ji (axvcb1597382@gmail.com), Tianran Liu (liutianran1121@gmail.com), Tonio Buonassisi (buonassisi@mit.edu)

**Abstract**

Scientific hypothesis generation is central to materials discovery, yet current approaches often emphasize either conceptual (idea-to-data) reasoning or data-driven (data-to-idea) analysis, rarely achieving an effective integration of both. Here, we present a generalizable active learning workflow that integrates top-down, theory-driven hypothesis generation, guided by a large language model. This is complemented by bottom-up, data-driven hypothesis testing through a root-cause association study. We demonstrate this approach through the design of equimolar quinary-cation two-dimensional perovskite, a chemically complex system with over 850,000 possible cation combinations. In the top-down component, the large language model drives closed-loop optimization by proposing candidates that are likely to achieve phase purity, leveraging domain knowledge and chain-of-thought reasoning. With each iteration, the model identifies an increasing number of near phase-pure compositions, sampling less than 0.004% of the design space. In parallel, the bottom-up association study identifies molecular features with statistically significant influences on phase purity. The integration of these approaches enables the convergence of conceptual and statistical hypotheses, leading to generalizable and rational design rules for phase-pure quinary-cation two-dimensional perovskites. As a proof of concept, we applied the optimized phase-pure quinary-cation two-dimensional perovskite film as a surface capping layer in perovskite solar cells, achieving good performance and stability. Our framework enables the development of interpretable and generalizable design rules that are applicable to a wide range of optimization processes within complex design spaces, providing a foundational strategy for rational, scalable, and efficient materials discovery.

## Introduction

Scientific hypotheses are testable statements that offer predictions or explanations for observed phenomena, forming the foundation of scientific exploration[1–3]. Hypotheses can be generated either at the beginning or at the end of a research cycle[4–6]. A researcher may first develop an idea, design experiments with a predicted outcome, and then collect experimental results that are either consistent or inconsistent with the hypothesis, resulting in a top-down, theory-driven approach[7]. Alternatively, rules (or hypotheses) can be inferred from examples and observations, forming a bottom-up, data-driven approach. Top-down and bottom-up science are not opposing approaches but rather complementary ones[8]. In material science, the integration of the two approaches enables researchers to explore intricate relationships among composition, structure, and properties, thereby providing a systematic framework to understand material characteristics and design novel materials[9–11].

Traditional trial-and-error methods, guided by top-down reasoning and expert intuition, are limited by their reliance on low-throughput, resource-intensive experimentation[12–14]. While these approaches have historically driven discovery, the manual extrapolation of design principles from small datasets hinders their generalizability across complex material systems such as halide perovskites. On the other hand, recent advances in artificial intelligence (AI) have enabled bottom-up, data-driven techniques through closed-loop experimental workflows, facilitated by Bayesian optimization (BO), for self-driven laboratories[15–20]. They support high-throughput experimentation, rapid generation and analysis of large datasets, and efficient optimization, which are promising to accelerate the materials screening and design process.

A key limitation of BO in closed-loop learning is its lack of interpretability, meaning that no explicit hypotheses or rules can be directly inferred, as it relies on surrogate models, such as Gaussian Processes (GPs), which function as black boxes and provide limited insight into the reasoning behind their suggestions. To solve this, it is often coupled with other machine learning algorithms, such as decision trees and regression models, and utilizes SHAP (SHapley Additive exPlanations) analysis to interpret model outputs by quantifying the contribution of each input feature to the prediction and ranking the most important ones. However, these approaches are limited by the risks of overfitting or underfitting in the selected machine learning models, and by the unreliability of SHAP values in interpreting models with strongly correlated features. These limitations restrict their ability to analyze causal relationships, extract actionable insights, or generate new hypotheses[21–24]. In addition, while BO is effective in many continuous settings, it faces challenges when applied to high-dimensional, discrete, or sparse search spaces, where traditional surrogate models and acquisition functions become inefficient[25–28]. Thus, an ideal closed-loop learning workflow should both provide explicit design rules and allow complex and high-dimensional search spaces.

Here we propose a novel workflow that is generalizable to a wide range of optimization processes in materials design. It integrates a closed-loop optimization framework driven by large language models (LLMs) as the top-down component, alongside a root-cause association study as the bottom-up component (**Fig. 1**). The LLM-driven closed-loop framework, which replaces BO, combines human experimental actions with LLM design actions: humans synthesize and characterize materials, while the LLM, prompted by natural language and tabular experimental data, analyzes preliminary results and proposes new candidates. For each candidate, the LLM generates a scientific hypothesis that provides top-down, abstract-level natural language rationales for potential material design rules toward the optimization target.

To complement the natural language rationales, we incorporate a root-cause association study, a method widely used in biomedical science for identifying genetic variants associated with human diseases[29–31]. This approach refines the abstract-level hypothesis into more specific, bottom-up feature-level hypotheses, which then allow for practical hypothesis testing through statistical methods. By integrating top-down, LLM-driven closed-loop experimentation with the statistical insights of bottom-up, data-driven analysis, our workflow enables the extraction of generalizable design rules, with broad applicability to diverse optimization processes in materials discovery.

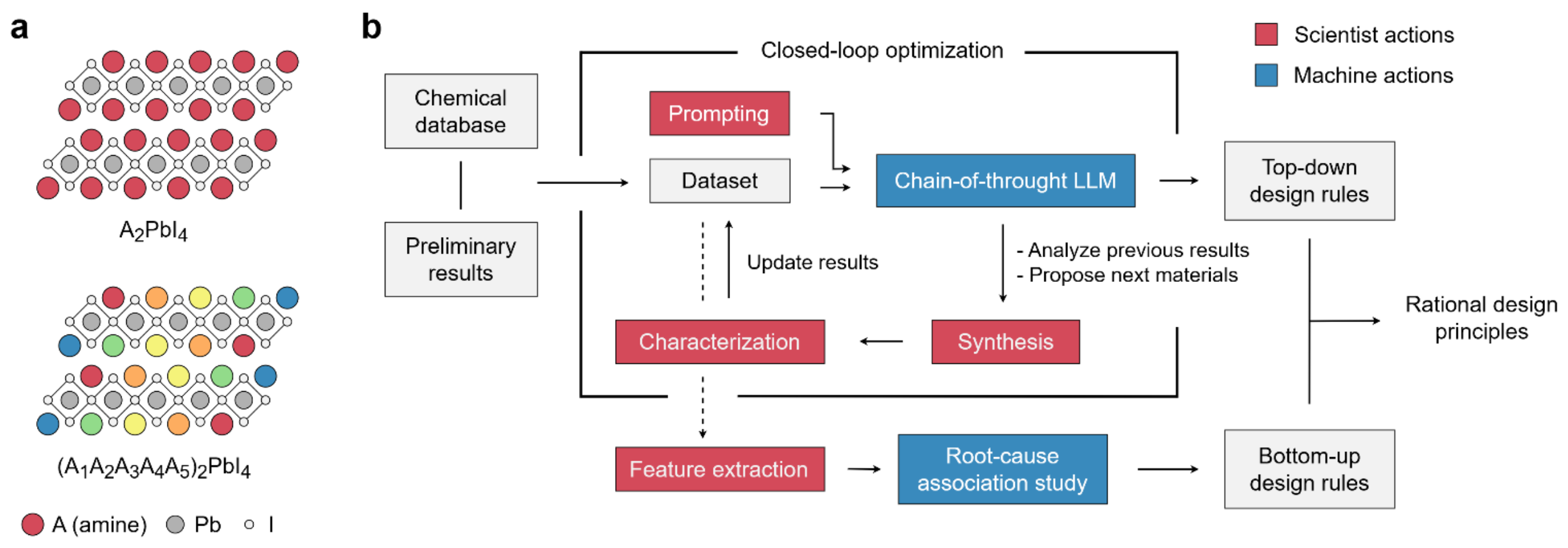

**Fig. 1. Overview of LLM-guided phase-pure equimolar quinary-cation 2D perovskite design strategy**. **a**, Schematic of layered perovskite structures showing single-cation 2D perovskites and quinary-cation 2D perovskites. **b**, Material design workflow that integrates closed-loop LLM-guided optimization and root-cause association study to enable top-down and bottom-up materials design principles.

## Results

In this study, we demonstrate our proposed workflow through designing novel two-dimensional (2D) lead halide perovskites through A-site organic cation engineering. The equimolar quinary-cation 2D perovskites, represented as $(A_1A_2A_3A_4A_5)_2PbI_4$ (**Fig. 1a**), have shown promise for optoelectronic applications due to their enhanced stability compared to their single-cation counterparts[32]. These materials leverage disordered organic moieties to achieve entropy stabilization through equimolar mixing of five distinct A-site organic cations, analogous to compositional strategies used in all-inorganic high-entropy materials[33]. However, achieving phase-pure quinary-cation 2D perovskite formation is nontrivial. Combinations of A-site cations can easily lead to 2D phase mixtures, which are frequently associated with reduced performance and compromised stability[32]. Designing phase-pure quinary-cation 2D perovskites remains a significant challenge due to the vast combinatorial complexity and the limited experimental exploration of the chemical space. For example, selecting five components from a pool of 42 A-site cation candidates that are available in our lab yields 850,668 unique 2D perovskite combinations, making systematic exploration through traditional experimental methods infeasible due to time and resource constraints. To date, only a few quinary-cation 2D perovskites with high phase purity have been synthesized successfully[32].

To guide the discovery of phase-pure quinary-cation 2D perovskites, our workflow integrates top-down active learning and reasoning via LLMs with bottom-up data-driven analysis via association studies (**Fig. 1b**). The workflow begins with an organic cation database comprising 42 ammonium cations for halide perovskites (**Supplementary Fig. 1**), screened through literature and curated by domain experts. To obtain preliminary experimental data, six combinations, each consisting of five randomly selected A-site cations from the database, were used to form quinary-cation 2D perovskite films. The samples were then characterized using X-ray diffraction (XRD) to evaluate their phase purity. The resulting preliminary dataset, together with the cation candidate list, is used to initiate the closed-loop optimization process. For each iteration, we used a commercial LLM, GPT-4o[34], to analyze the existing data and propose new rounds of candidates with proposed hypotheses in natural language. These LLM-suggested combinations were then synthesized, characterized, and evaluated for phase purity by human researchers. The new results were fed back into the dataset to initiate the next round of learning with the LLM, which then suggested new candidates and enabled the continued iterative refinement of the design space.

For each candidate, the molecular features (to be introduced later) of five selected organic cations were compiled into a one-dimensional (1D) feature array. This array, together with phase purity of the perovskite sample, was used for statistical analysis to identify features that were statistically significant in influencing the phase purity of quinary-cation 2D perovskites. It correlates molecular structures with desired properties, contributing to the development of bottom-up, data-driven design rules. Then, the top-down conceptual hypotheses are aligned with bottom-up statistical insights through semantic similarity matching via LLM, forming concrete design principles for phase-pure quinary-cation 2D perovskites within a closed-loop, data-driven framework.

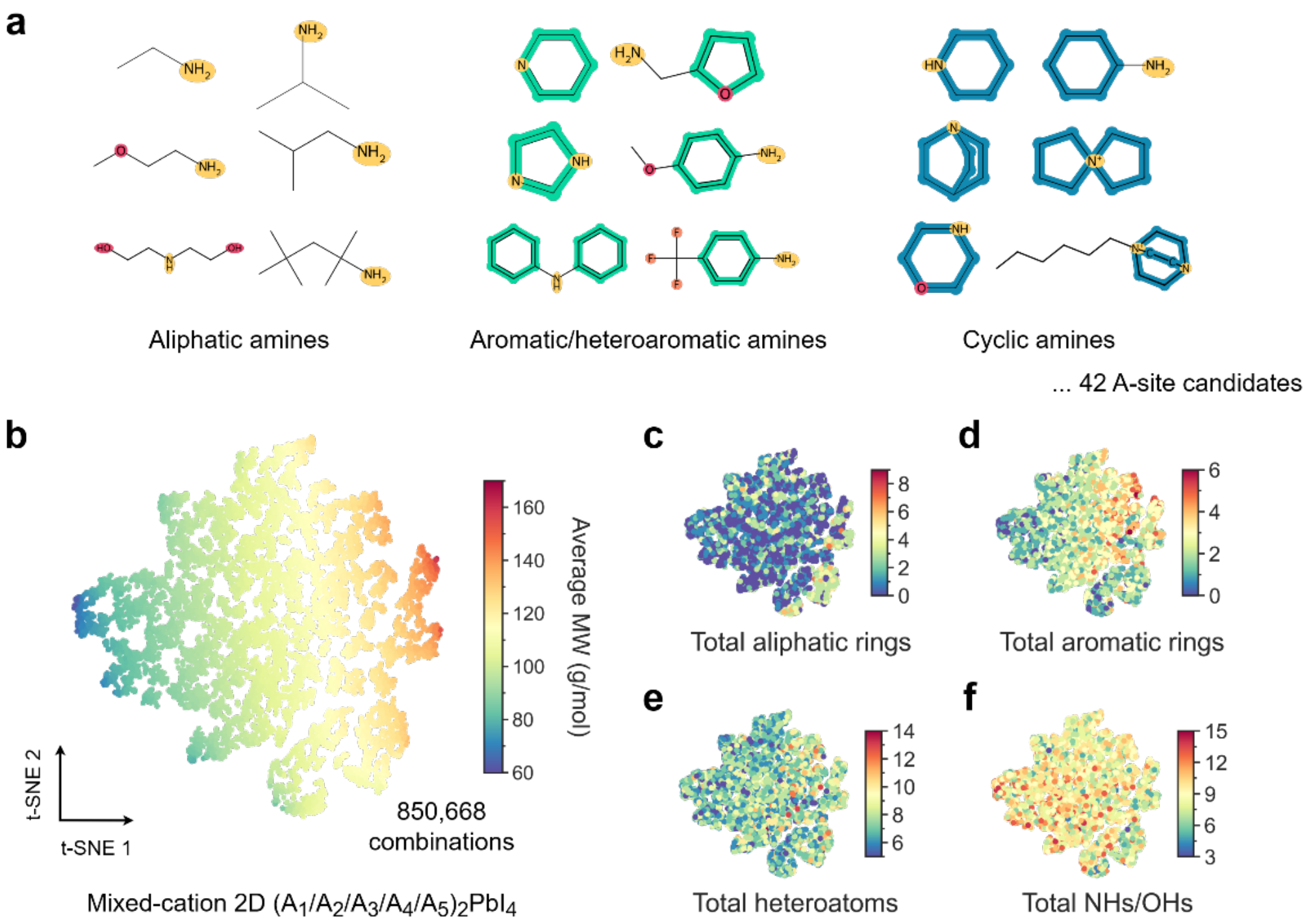

**Fig. 2. Chemical search space for equimolar quinary-cation 2D perovskite**. **a**, Molecular structures of representative A-site organic cations in our chemical database, categorized as aliphatic amines, aromatic/heteroaromatic amines, and cyclic amines, are depicted in their neutral form for simplicity. **b**-**f**, t-SNE visualization of the chemical space for quinary-cation 2D perovskites containing five A-site cations, color-coded by (**b**) average molecular weight (MW), (**c**) total aliphatic ring count, (**d**) total aromatic ring count, (**e**) total heteroatom count, and (**f**) total count of NH/OH functional groups.

We first defined the chemical search space by selecting 42 A-site cations that have the potential to form 2D perovskites and are available for purchase, including aliphatic amines, aromatic/heteroaromatic amines, and cyclic amines (**Fig. 2a**). To visualize and understand this vast chemical space, we employ t-distributed stochastic neighbor embedding (t-SNE) to project the search space into two dimensions and highlight the variation of different features across it, including average molecular weights (MW), aliphatic/aromatic ring counts, heteroatoms, and functional groups (**Fig. 2**, **b-f**) [35].

We then defined the learning objective as maximizing the phase purity of the quinary-cation candidates. However, directly quantifying phase purity through XRD results is challenging for perovskite thin-film samples due to preferred orientation and missing reference patterns, as these quinary-cation materials have never been synthesized or characterized using powder XRD. Therefore, we created an arbitrary proxy metric for phase purity as one minus the ratio of the second-highest XRD peak intensity (the secondary phase) to the highest peak intensity (the primary phase). The proxy phase purity metric is manually extracted from XRD patterns for each composition to avoid mistakenly identifying impurities, such as peaks from higher-order reflections of the primary phase.

We started the closed-loop optimization by preparing a preliminary dataset in Round 1, where six candidates were randomly selected and synthesized by human scientists, and their phase purity metric was calculated from the XRD measurements. For each candidate, the proposed five ammonium iodides were stoichiometrically mixed with $PbI_2$ in dimethylformamide (DMF) to prepare the $(A_1A_2A_3A_4A_5)_2PbI_4$ precursor solution, which was used to spin coat thin films of quinary-cation perovskites. In subsequent rounds (Rounds 2–5), prior experimental results were analyzed by the LLM, and the next six candidates were proposed with accompanying rationales (see Method for prompting). In total, 30 cation combinations were explored (**Fig. 3a**), and the raw diffraction patterns of the synthesized 2D perovskites are demonstrated in **Supplementary Fig. 2-6**.

To quantify the diversity and novelty of the proposed candidates across the search space, we calculated both the exploration and the combination novelty scores. The exploration score was determined by computing the smallest City Block (Manhattan) distance between the one-dimensional (1D) feature array of each new candidate and the previous-round candidates. A higher score indicates a greater difference (or more diversity) in the combination of new organic cation structures compared to earlier candidates. Combination novelty is defined as the maximum number of overlapping cations between new and previous round candidates (e.g., a score of 3 indicates that up to 3 out of 5 cations were used in one or more past combinations).

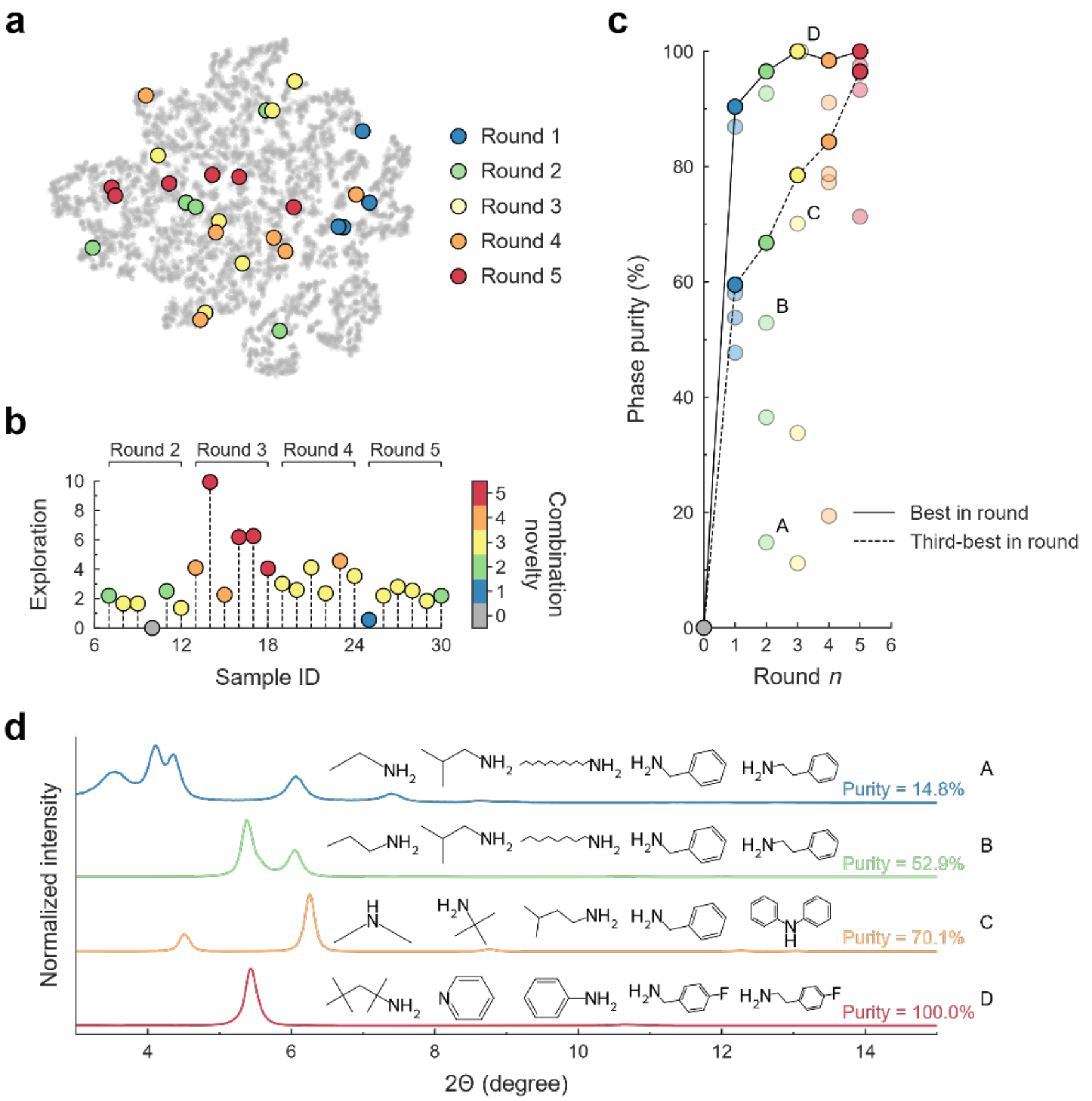

**Fig. 3. Iterative optimization of quinary-cation 2D perovskites using an LLM-guided closed-loop workflow. a**, Distribution of synthesized combinations across the 5 rounds of the closed-loop optimization process visualized by t-SNE. **b**, Exploration scores and combination novelty of each synthesized sample from Rounds 2–5, with color indicating the degree of novelty relative to prior rounds. **c**, Phase purities (%) of all samples synthesized for each round. Solid and dashed lines connect the top-best and the third-best samples. **d**, Representative XRD patterns of 4 selected synthesized quinary-cation

$(A_1A_2A_3A_4A_5)_2PbI_4$ 2D perovskites (A–D) with increasing phase purity, along with their cation molecular structures and measured phase purity values.

We plotted the exploration score against the sample ID and color-coded the samples according to their combination novelty (**Fig. 3b**). The graph illustrates how exploration (trying new structural and molecular combinations) and exploitation (sticking to similar combinations to improve purity) vary across different rounds. In particular, Round 3 tends to be more explorative, with an average exploration score of 5.46 and an average combination novelty score of 4.67, as we deliberately encouraged exploration through prompts (see **Supplementary Text**). The later rounds (Rounds 4 and 5) exhibit relatively lower average exploration scores (3.36 and 2.02) and combination novelty scores (3.16 and 2.50), indicating a shift toward exploitation by the LLM.

The phase purity scores of proposed candidates in each round are plotted in **Fig. 3c**, with the solid line representing the top-best candidate in each round and the dashed line indicating the third-best. We observe that Round 1, which relied on random human selection, yielded a top-best purity of 90.4% and a third-best purity of only 59.5%. In Rounds 2-5, guided by LLM exploration, the phase purity of the top-best candidate plateaued above 98% and approached 100% as early as Round 3, while the third-best purity steadily increased, reaching 98% in Round 5. The results show that the LLM proposed more high-purity candidates as the learning cycles progressed, reflecting its ability to rapidly and effectively screen the chemical search space for optimal solutions.

These favorable outcomes were achieved by experimentally sampling less than 0.004% of the total 850,668 possible combinations, showing the efficiency of the LLM-guided workflow. We infer that this effectiveness results from the LLM's prior domain knowledge, which enables it to recognize structure-property relationships and efficiently guide composition selections across complex design spaces. Overall, these results validate the potential of LLM-driven closed-loop optimization workflow to deliver precise and consistent predictions efficiently in high-dimensional sparse design spaces that are otherwise impractical to explore through BO methods. **Fig. 3d** presents representative XRD patterns for quinary-cation 2D perovskite compositions and their corresponding A-site cation chemical structures, illustrating the variation in phase purity from 14.8% to 100%. Low phase-purity 2D perovskites were identified by multiple diffraction peaks corresponding to single-component 2D perovskite structures, while high phase-purity 2D perovskites exhibited clean and well-defined peaks[32,33,36].

In each round, we prompted the LLM to provide rationales for each candidate combination, leveraging its chain-of-thought (CoT) reasoning[37] and its ability to express conceptual hypotheses through natural language[38]. The word cloud[39] results of rationales highlight several prominent themes related to high phase purity, including "aromatic", "packing", "steric", and "fluorinated" (**Fig. 4a**). We further categorized these conceptual hypotheses into chemical and structural categories with three domain experts (average Cohen's Kappa score = 0.793; **Supplementary Table 1**)[40], and summarized their word frequencies in **Fig. 4b**. Categories such as "aromatic", "heteroatom/functional", "steric/bulky", "bonding/electronic/force", and "aliphatic/alkyl" appeared most frequently in the proposed hypotheses.

While the top-down hypotheses from LLM are valuable for guiding materials design, they are conceptual and qualitative, lacking the specificity required to establish generalizable and robust design principles. To complement these, we performed a bottom-up association study to quantitatively assess which combinations of molecular features influence phase purity. Association studies are a well-established approach in biomedical science to identify statistical relationships between genetic variants and specific traits or diseases, helping to uncover potential biological mechanisms and risk factors[41,42]. Inspired by this framework, we first calculated 15 molecular and structural properties (i.e., descriptors) for each A-site cation, including molecular weight (MolWt), fraction of $SP^3$-hybridized carbons (FractionCSP3), number of NH or OH groups, number of aliphatic or aromatic carbocycles, heterocycles, or rings, number of hydrogen bond donors or acceptors, number of heteroatoms, number of rotatable bonds, and total ring count. We then hypothesized that the mean or variance of one or more descriptors across combinations of five cations could affect phase purity, and performed regression to test these bottom-up hypotheses (**Supplementary Fig. 7**).

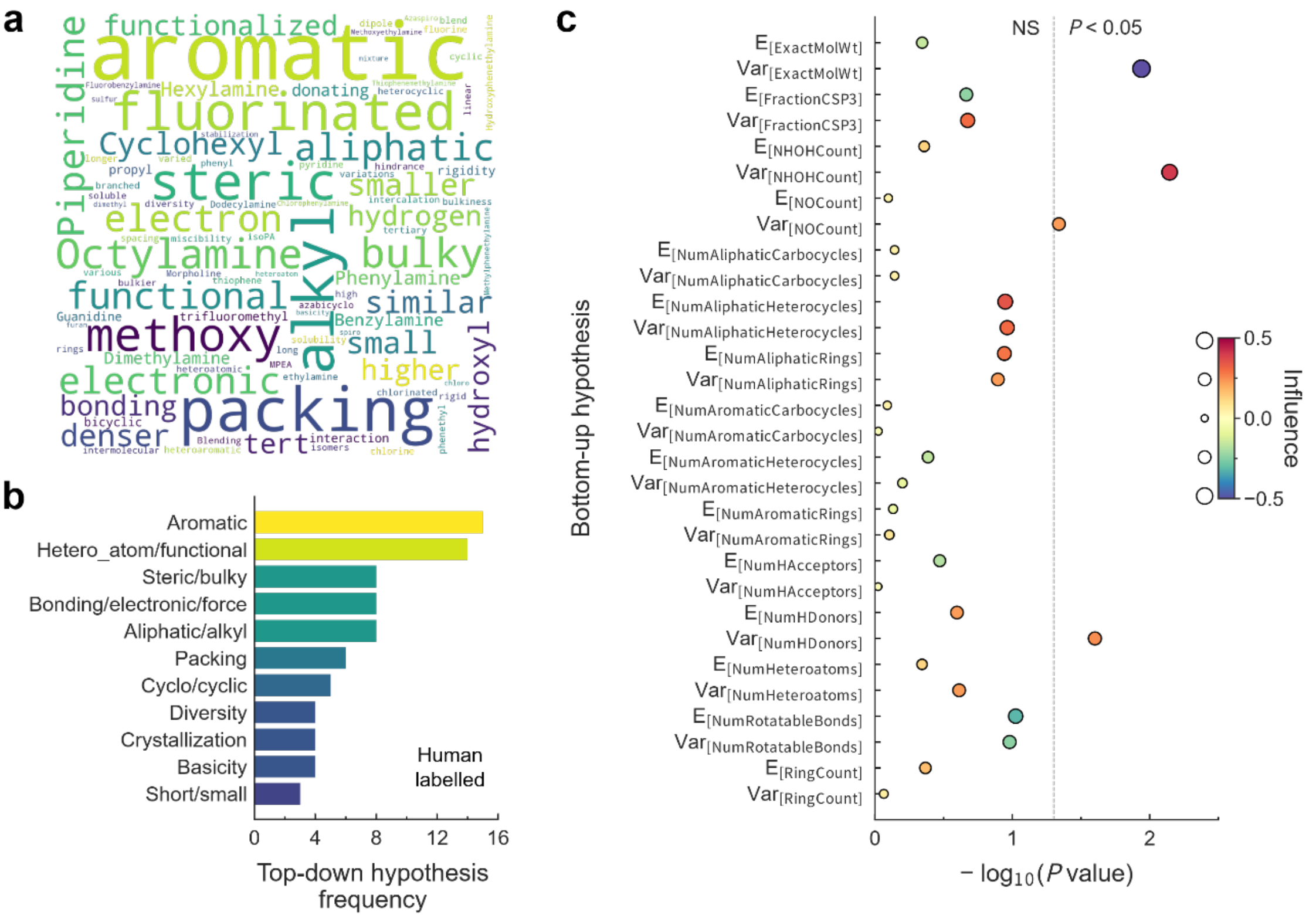

**Fig. 4. Top-down conceptual hypothesis and bottom-up statistical analysis that reveal key molecular features influencing phase purity. a**, Word cloud summarizing top-down hypotheses generated by the LLM during closed-loop optimization, which shows key features of five A-site cations in each combination hypothesized to influence phase purity of quinary-cation 2D perovskites. Font size corresponds to the frequency with which each descriptor was mentioned by the LLM. **b**, Bar plot showing the frequency of top-down conceptual hypotheses generated by the LLM, categorized by chemical or structural features of A-site cations via human experts. **c**, Bottom-up statistical analysis that uses root-cause association study to identify the influence of molecular features extracted from five A-site cations in each combination on phase purity of the quinary-cation 2D perovskites. A plot of -$\log_{10}$($p$-value) against molecular features of each combination, depicting their influence (effect direction and size) and significance ($p$-value) on phase purity. Each dot represents an individual descriptor,

with color indicating the direction of influence (red: positive correlation; blue: negative correlation), and size reflecting the magnitude of influence (larger size means stronger effect). The analysis was performed using Ordinary Least Squares regression (**Methods** and **Supplementary Fig. 7**). "NS" denotes no significance, while a $p$-value < 0.05 denotes statistical significance.

**Fig. 4c** visualizes the $-\log_{10}(P \text{ value})$ and influence for each bottom-up hypothesis in a Manhattan plot. The p-value is used for hypothesis testing against the null hypothesis, and the influence value quantifies the effect, while its sign (positive or negative) denotes whether the feature promotes or suppresses phase purity. We identified several features with statistical significance on phase purity. These include the variance in exact molecular weights, the variance in the number of NH or OH groups, the variance in the number of nitrogen and oxygen atoms, and the variance in the number of hydrogen bond donors. Notably, the variance of exact molecular weight across the five cations exhibited a strong negative correlation with phase purity ($P = 0.011$; influence = -0.53), while the variance in NH or OH group counts across the five cations showed a statistically significant positive correlation ($P = 0.0072$; influence = 0.40) [43]. These results suggest that our bottom-up approach is able to quantitatively identify key molecular factors that influence phase purity, establishing data-driven design rules for achieving high-purity quinary-cation 2D perovskites.

We noted that many of the bottom-up hypotheses are correlated with each other. We demonstrated the intercorrelations among molecular and structural features and their relationships to phase purity using a Pearson correlation matrix (**Fig. 5a**). We found that the three previously identified statistically significant hypotheses (the variance in the number of NH or OH groups, the variance in the number of nitrogen and oxygen atoms, and the variance in the number of hydrogen bond donors) are strongly correlated with each other, which indicates that these data-driven findings can be further consolidated into one central hypothesis.

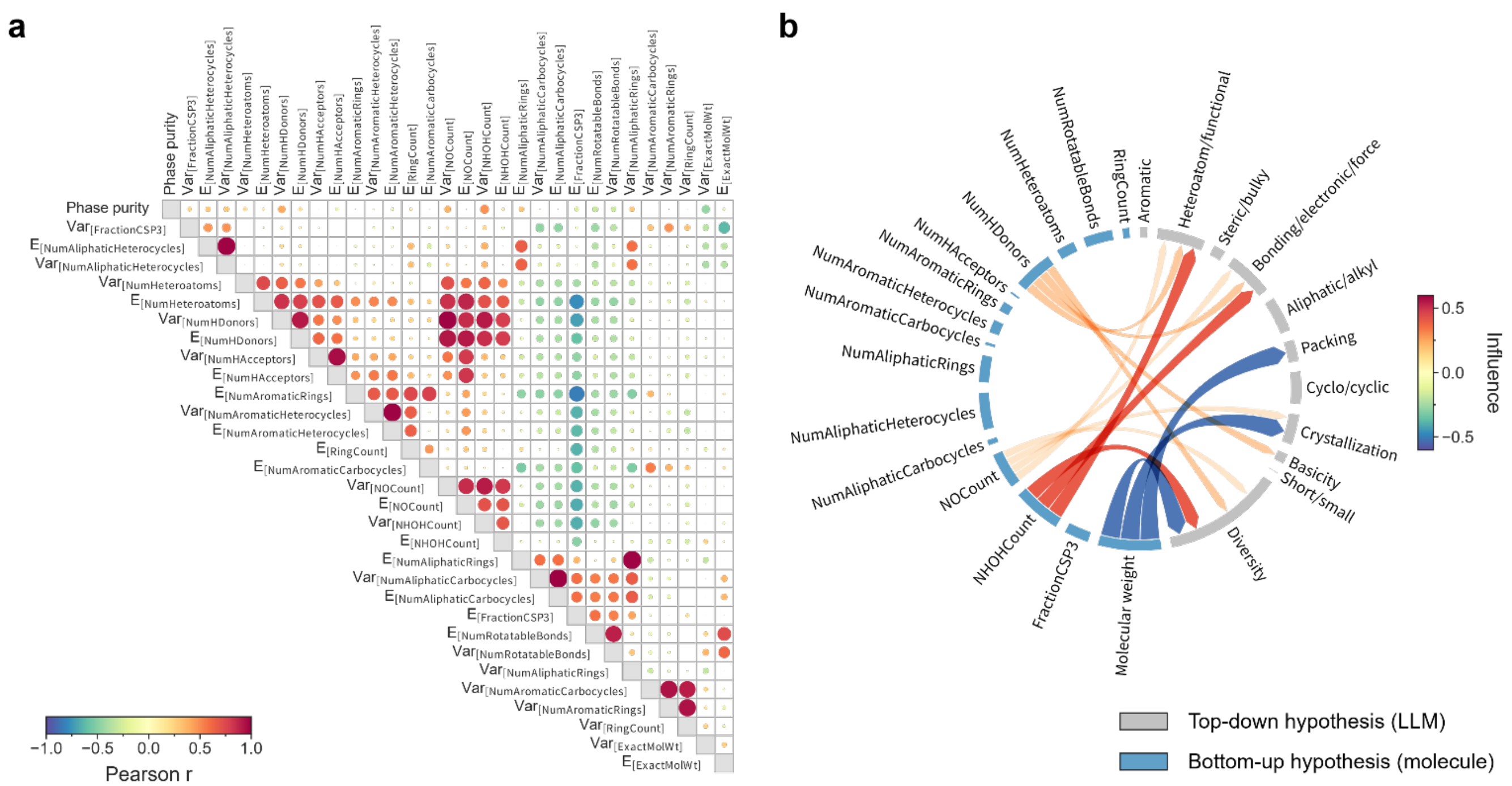

**Fig. 5. Analysis of quinary-cation molecular features influencing phase purity. a**, Pearson correlation matrix displaying pairwise correlations (*r*) among molecular features extracted from the five A-site cations in each composition and phase purity. For clarity, the feature order was grouped and sorted using flat hierarchical clustering to place molecular features with similar correlation patterns adjacent to each other. Each circle represents the strength and direction of correlation between two features (red: positive, blue: negative), with size and color intensity corresponding to the magnitude of *r*. **b**, Integration of top-down (gray) and bottom-up (blue) hypotheses on the influence of feature variances on phase purity. The thickness of a connecting line represents its magnitude of influence (i.e., the gradient of regression, **Supplementary Fig. 7**). The thickness of a hypothesis corresponds to the sum of the total influences of its connecting lines. Each line is then colored to represent the direction of the influence, with transparency indicating significance: opaque for $P < 0.01$ and fully transparent for $P > 0.1$.

To obtain the rational design rules, we used semantic similarity matching to connect the top-down and bottom-up hypotheses (**Supplementary Table 2**), where each bottom-up hypothesis is linked to one or more top-down hypotheses through a chord diagram (**Fig. 5b** and **Supplementary Fig. 8**). We excluded connections with high $P$ values ($P > 0.1$) by rendering them fully transparent. We further prompted LLM models to summarize the available connections, focusing on molecular weight variance and three previously identified, strongly correlated hypotheses (**Methods**). This led to two final design rules: (i) low variance in molecular weight to promote compatible packing and synchronized crystallization, and (ii) high variance in hydrogen-bonding capacity and heteroatom-containing functional groups to encourage complementary intermolecular interactions that stabilize a single perovskite phase.

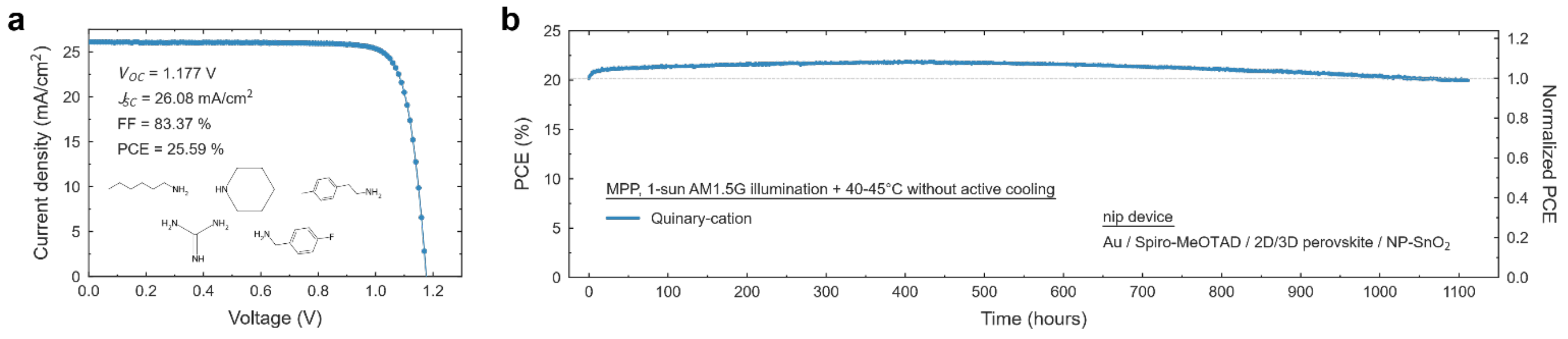

**Fig. 6. Efficiency and stability of 2D/3D bilayer heterostructured PSCs based on quinary-cation 2D perovskite capping layer. a**, Current density-voltage curve of PSCs based on representative high-purity 2D capping layer (Round 5 candidate 3). **b**, Maximum power point tracking under 1-sun AM1.5G illumination (UV included) of the device at 40°C to 45°C in a nitrogen atmosphere.

To demonstrate the practical applicability of the quinary-cation 2D perovskites identified through our framework, we integrated them into 2D/3D bilayer heterostructured perovskite solar cells (PSCs) (**Fig. 6**). We selected the third composition from round 5 ("5-3"), which exhibited nearly 100% phase purity, and the champion device achieved a champion power-conversion efficiency (PCE) of 25.59% (**Fig. 6a**). We attribute the superior performance of the quinary-cation device to the highly phase-pure 2D interlayer and the synergistic effect of multiple A-site cations interacting with diverse surface defects, enabling more effective passivation and reduced electronic disorder[32]. It is noteworthy that our device uses rare cations which are not commonly found in the literature, such as guanidinium. We speculate guanidinium's delocalized positive charge may help screen local charge fluctuations from dipolar species (e.g., fluorinated and cyclic amines), while also passivating undercoordinated Pb sites through multidentate binding[44]. This is consistent with the results from association studies on molecular features calculated from first principles (**Supplementary Fig. 9**), which shows that both the mean and variance of the electrostatic potentials contribute positively to phase purity ($P$ = 0.018, influence = 0.48; $P$ = 0.00016, influence = 0.59, respectively). We posit that identification of this unique combination of uncommonly used quinary cations would have been challenging without our integrative framework, if relying solely on ordinary experimental techniques and trial-and-error approaches.

Although the optimization approach used a single proxy variable for phase purity, we nevertheless achieved state-of-the-art performance and promising stability. Our device showed excellent operational photostability, retaining ~99% of its initial PCE (or ~91% of its peak PCE) after over 1100 hours under continuous 1-sun illumination at 40-45°C in nitrogen (**Fig. 6b**), among the best reported for n-i-p structured PSCs[45]. We attribute the device stability to the high phase purity and entropy-driven structural resilience of the quinary-cation composition[32,36]. Altogether, the device results highlight the potential of our framework for rational materials design to predict quinary-cation 2D perovskite compositions. These results echo prior findings in related systems. For instance, similar synergies have been observed in nanocrystal ligand engineering, where mixed-ligand shells enhance colloidal and thermal stability compared to single-ligand systems[46]. This body of evidence points to an emerging principle: well-chosen mixtures of molecules can outperform any single component when it comes to stabilizing complex, soft-matter interfaces, suggesting the potential of our framework to be extended to other complex systems for optimization.

Together, this integrative framework offers a holistic understanding of the molecular design space, highlighting how top-down and bottom-up hypotheses converge to inform rational design rules for phase-pure quinary-cation 2D perovskites. This convergence demonstrates the complementarity of the two approaches: bottom-up hypotheses provide statistically robust, data-driven validation, while top-down hypotheses offer conceptual clarity and interpretability, helping to prevent data fishing/spurious correlations by grounding the analysis in domain knowledge and scientific reasoning. Their integration ensures that the resulting design rules are not only explainable and rational, but also empirically supported and generalizable.

**Discussion**

We develop a generalizable active learning workflow that integrates top-down theory-driven reasoning via LLM-guided closed-loop experimentation with bottom-up data-driven analysis based on association study. The LLM-guided materials design, even with sparse data, can yield complex formulations with synergistic, emergent properties. Demonstrated through a case study on the rational design of phase-pure quinary-cation 2D perovskites, this framework rapidly and effectively navigates a chemically complex system with over 850,000 possible A-site combinations to identify high-purity composition candidates with minimal experimental effort. The LLM-driven optimization efficiently proposes promising compositions by leveraging prior scientific knowledge, while the association study provides quantitative insights into the molecular features most critical to phase behavior, together enabling the formulation of rational design rules for phase-pure quinary-cation 2D perovskites.

This integration of top-down and bottom-up reasoning is complementary and addresses the limitations of each individual approach. While top-down hypotheses offer interpretability and mechanistic insights, they often lack statistical rigor; conversely, bottom-up analyses provide quantitative validation but may lead to spurious correlations without clear physical justification. Their integration enables principled hypothesis generation and empirical validation within a unified framework. As a result, this hybrid strategy yields statistically grounded, interpretable, and generalizable design rules, laying the foundation for rational, scalable, and efficient materials discovery.

While we demonstrated the use of five cations to showcase a capability that surpasses traditional trial-and-error methods, determining the ideal number of components remains a challenge. This optimal number will likely depend on the number of independent functions required, such as moisture resistance, defect passivation, and interfacial adhesion. Another limitation involves the quality and functional specialization of those molecules. It remains unclear whether it is more effective to use a single molecule that offers multiple desirable traits or several molecules, each targeting a specific function. While a single-molecule approach may simplify processing and reduce entropy-driven disorder, a mixed-molecule strategy could enable more precise tuning of material properties. These trade-offs underscore the need for further investigation.

Beyond the specific material system studied here, our framework offers a scalable and adaptable approach for accelerating materials optimization across diverse applications. By bridging language-based reasoning with data-driven analysis, it enables automated, hypothesis-driven, and interpretable materials design. With the incorporation of additional optimization objectives and associated experimental data, e.g., stability data from an accelerated stress test, this framework could offer a route to design PSCs with a constellation of industrially relevant properties. Looking ahead, integrating automated synthesis and high-throughput characterization could further boost the workflow's speed and efficiency, advancing toward fully autonomous discovery pipelines from hypothesis to validation. Our work is a first step, but an important one, toward the broader vision of AI-guided discovery of materials with engineered property constellations for real-world applications in photovoltaics and beyond.

## Methods

### *Materials*

All materials were purchased from Sigma-Aldrich, unless otherwise stated. All ammonium iodides were purchased from Greatcell Solar Materials. $SnO_2$ colloidal solution was purchased from Alfa Aesar

Chemicals. 2,2',7,7'-tetrakis(N,N -di-p methoxyphenylamino)-9,9'-spirobifluorene (Spiro-OMeTAD) was purchased from Luminescence Technology Corp. N-Propyl-3-methylpyridinium bis(trifluoromethylsulfonyl)imide was purchased from Strem Chemicals. All chemicals were used directly as received without further purification.

*Quinary-cation 2D perovskite synthesis*

Five ammonium iodides were stoichiometrically mixed with $PbI_2$ in N,N-dimethylformamide (DMF) to prepare the $(A_1A_2A_3A_4A_5)_2PbI_4$ precursor solution. Specifically, 0.5 M of each ammonium iodide (five total) and 0.25 M $PbI_2$ were dissolved in DMF to form an equimolar quinary-cation 2D perovskite solution. The mixture was stirred at room temperature until fully dissolved. Perovskite films were fabricated by spin-coating the solution at 3000 rpm for 30 seconds, followed by thermal annealing at 100 °C for 10 minutes. All processing steps were carried out in a nitrogen-filled glovebox maintained at 25 °C.

*Perovskite solar cell fabrication*

FTO substrates were cleaned by ultrasonication in Hellmanex, deionized water, acetone, and isopropanol for 10 min each. For chemical bath deposition of $SnO_2$, 137.5 mg of $SnCl_2 \cdot 2H_2O$, 625 mg of urea, 625 μL of hydrochloric acid, and 12.5 μL of thioglycolic acid were mixed in 50 mL deionized water. The front electrode of the FTO substrates was taped using Kapton tape before $SnO_2$ deposition. FTO substrates were then loaded into a Hellendahl glass reaction vessel (volume ~170 mL) and heated at 65 °C for 14 hours in a water bath. Subsequently, the $SnO_2$/FTO substrates were cleaned by ultrasonication in deionized water and isopropanol for 5 min each. Finally, the substrates were then annealed for 1 hour at 170 °C in an ambient environment. For nanoparticle $SnO_2$, the $SnO_2$ colloidal solution was diluted in deionized water in 1:6.5 ratio. FTO substrates were treated by oxygen plasma for 10 mins. The diluted $SnO_2$ was then spun at 3000 rpm for 30 secs. The $SnO_2$/FTO substrates were then annealed for 30 mins in an ambient environment at 150 °C.

For passivation of the buried interface, the $SnO_2$/FTO substrates were treated by oxygen plasma for 10 mins. 10 mM KCl in deionized water was spun at 3000 rpm for 30 secs, followed by annealing for 10 mins at 100 °C. The FTO/$SnO_2$ substrates were treated again by oxygen plasma for 10 mins before perovskite deposition. The perovskite precursor solution was prepared by mixing 1.4 M FAI, 0.5 M MACl, 0.013 M $MAPbBr_3$, and 1.53 M $PbI_2$ in a mixed DMF/DMSO solvent with volume ratio 8:1. The solution was stirred at room temperature for at least 3 hours with a magnetic stirrer before use. The solution was filtered with a 0.22 μm PTFE filter before use. The 3D perovskite was deposited via spin coating at 500 rpm for 7 secs, then 1000 rpm for 8 secs, then 5000 rpm for 30 secs. 10 secs into the 5000 rpm stage, 700 μL of diethyl ether antisolvent was dropped on the wet film. The films were then annealed successively first at 100 °C for 40 mins, then 150 °C for 5 mins. All processing steps for the perovskite films are conducted in a nitrogen glovebox at a temperature of 25 °C. For quinary-cation 2D perovskite deposition, 12 mM of the quinary-cation solution was dissolved in isopropanol and deposited at 4000 rpm for 30 secs. The spiro-OMeTAD hole transporting layer solution was prepared by mixing 90.0 mg spiro-OMeTAD, 9.0 mg of $spiro^{2+}(TFSI^{-})_2$, and 3.17 μL of N-Propyl-3-methylpyridinium bis(trifluoromethylsulfonyl)imide in 1 mL of chlorobenzene. The solution was filtered with a 0.22 μm PTFE filter before use. The solution was deposited via spin coating at 4000 rpm for 20 secs. Lastly, 100 nm gold electrode was deposited via thermal vacuum evaporation at a rate of 0.5 A $s^{-1}$ for the first 10 nm, then 1.0 A $s^{-1}$ for the rest 90 nm.

*Device stability testing*

The stability test was conducted by tracking the maximum power point (MPP) under full-spectrum 1-sun AM 1.5G illumination using a G2V Optics Pico LED solar simulator, without wavelength filtering (e.g. no UV cutoff), in a nitrogen atmosphere. Without active heating or cooling, the measured device temperature was 40-45 °C. The device for the stability test used nanoparticle $SnO_2$ as the electron transport layer.

*X-ray Diffraction*

X-ray diffraction patterns were collected by a Rigaku SmartLab X-ray diffractometer with a Cu-Kα source under the ambient environment. The grazing incidence angle was 0.5°. The collecting 2θ angle range was 3-15°, with a step size of 0.02° and a scanning speed of 6°/min.

*Prompt Design*

To prompt the LLM to suggest new composition combinations for the next round, we used a zero-shot format without providing any example inputs (**Supplementary Text**). In each round, the prompt basically included a brief introduction to equimolar quinary-cation 2D perovskites, a definition of phase purity, a request for the LLM to propose new combinations based on previous phase purity results, a table listing all ligand candidates, and a second table summarizing the phase purity outcomes from earlier rounds. Some of the initially suggested candidates were found to be insoluble in DMF, which hindered the solution deposition of uniform films. To address this, an additional prompt was provided to inform the LLM of the solubility issue and request a revised set of candidate compositions.

**Example query**: I am going to make equimolar quinary-cation 2D perovskites based on A site. The composition will be $A_2PbI_4$, and I am targeting 5 ligands for A site. I have A site iodide based on A site halides in the form "ligand candidates". I define phase purity by using one minus the ratio of the highest

XRD peak value to the 2nd highest XRD peak value. In my 0th trial, I have tried six combinations and measured their XRD, and obtained their phase purity. Please read and suggest to me six more combinations for high-purity equimolar quinary-cation 2D perovskites with the rationale for each combination.combinations. I need the rationale for each combination.

**Answer**: The first file contains data from your initial trial of six combinations of A-site ligands, including their XRD peak values and phase purity. The second file lists various ligand candidates for potential A-site cations, providing details such as the halide type, cation mass, charge, and additional identifiers.

Based on the provided data and the goal of optimizing phase purity for quinary-cation perovskites, I'll suggest six new ligand combinations from the candidate list. I'll focus on maintaining diversity in chemical structure and size while selecting ligands with a balanced range of cation mass and charge, aiming to stabilize the lattice and improve phase purity.

Let me review the candidate list and suggest the new combinations with rationale.

To suggest six new combinations, I'll aim for a balanced selection of ligands based on their size (cation mass), structure (linear vs branched chains), and charge. This diversity may contribute to improved phase stability and higher phase purity. Here's a proposed rationale and six combinations:

1. Ethylamine iodide (EA), iso-Propylamine iodide (isoPA), Cyclohexylamine iodide (CHA), Phenethylamine iodide (PEA), n-Butylamine iodide (BA):
    - Rationale: A mix of small (EA, isoPA) and bulkier ligands (CHA, PEA) helps manage packing in the crystal lattice. Cyclohexyl and phenethyl groups are known to improve structural integrity in similar systems.
2. n-Propylamine iodide (PA), iso-Butylamine iodide (isoBA), n-Octylamine iodide (OA), Phenethylamine iodide (PEA), Benzylamine iodide (BZA):

- Rationale: Introducing a slightly longer chain (n-Octylamine) to stabilize intercalation, along with a blend of linear and bulky ligands, could balance both phase stability and crystallization.
3. iso-Butylamine iodide (isoBA), iso-Propylamine iodide (isoPA), Phenethylamine iodide (PEA), n-Octylamine iodide (OA), Benzylamine iodide (BZA):
    - Rationale: Combining multiple isomers with more branched alkyl groups could lead to denser packing, improving phase purity and reducing phase segregation.
4. n-Octylamine iodide (OA), Cyclohexylamine iodide (CHA), n-Butylamine iodide (BA), Phenethylamine iodide (PEA), 4-Methylphenethylamine iodide (4-MPEA):
    - Rationale: A combination of aliphatic, cyclohexyl, and aromatic amines could stabilize various packing structures, improving XRD peak purity.
5. n-Butylamine iodide (BA), Benzylamine iodide (BZA), Phenethylamine iodide (PEA), iso-Propylamine iodide (isoPA), 4-Methylphenethylamine iodide (4-MPEA):
    - Rationale: Blending aromatic (BZA, PEA, 4-MPEA) and alkyl ligands might enhance crystal integrity and mitigate halide segregation.
6. Ethylamine iodide (EA), n-Dodecylamine iodide (DA), iso-Butylamine iodide (isoBA), Phenethylamine iodide (PEA), Benzylamine iodide (BZA):
    - Rationale: The long-chain n-Dodecylamine may help with interlayer spacing while smaller ethylamine promotes denser packing, improving stability.

These combinations aim to balance steric effects and charge compatibility for improved phase purity. You can try these in your next trials.

*Association study*

In this study, we employed Ordinary Least Squares (OLS) regression to investigate the association between the variables (combinations of molecular features) and the outcome (phase purity). Each variable represents the mean or variance of a specific molecular feature calculated across five cation molecules for all the candidates. Prior to analysis, all variables were normalized to the range [0, 1], and multicollinearity among predictors was assessed by calculating the Pearson correlation matrix (**Fig. 5a**). In our case, variables with high collinearity were not removed or combined. The regression coefficients, standard errors, and p-values were estimated for each predictor, with significance determined at a nominal threshold of $p < 0.05$. Detailed regression plots with 95% confidence intervals are provided in **Supplementary Fig. 7**.

*Semantic similarity matching via LLM*

Traditional semantic similarity methods, which calculate cosine similarities on word embeddings, have limitations in incorporating factual and commonsense knowledge, and often perform poorly in specialized domains. In this work, we used GPT-4o to automatically connect bottom-up and top-down hypotheses, leveraging its ability to infer relationships that go beyond surface-level token similarity.

**Example query**: Match "variance of molecular weight" semantically with one or more of the following hypotheses: ['Aromatic', 'Hetero_atom/functional', 'Steric/bulky', 'Bonding/electronic/force', 'Aliphatic/alkyl', 'Packing', 'Cyclo/cyclic', 'Diversity', 'Crystallization', 'Basicity', 'Short/small']

**Answer**: The phrase "variance of molecular weight" refers to the range or diversity of molecular weights in a set of molecules. This concept aligns best with: *Diversity*. Reasoning: Variance in molecular weight directly reflects molecular diversity. A broad distribution of molecular weights typically indicates a chemically diverse set of compounds. Possible weak connections (less direct): *Packing*: Molecular weight can influence packing in solids, especially in polymers or crystal lattices, but it's an indirect link.

*Crystallization*: Variations in molecular weight can affect crystallization behavior, but again, this is an indirect or system-dependent effect. Not a semantic match: *Aromatic*, *Hetero_atom/functional*, *Steric/bulky*, *Bonding/electronic/force*, *Aliphatic/alkyl*, *Cyclo/cyclic*, *Basicity*, *Short/small* — These refer to structural or chemical features, not variance or distribution in molecular size or mass. Final match: Primary: *Diversity*. Secondary (context-dependent): *Packing*, *Crystallization* (possible but weaker links).

**Data availability**

The data supporting the findings of the study are available within the article and the Supplementary Information.

**Code availability**

All the LLM prompts used in this study are provided in the Method section.

**Acknowledgments**

We thank Otto Beall, Vinn Nguyen, Leala Nakagawa, Emilia Szczepaniak, Ronaldo Lee, and John Dranchak for help in sample preparation. K.J. and F.S. thank First Solar, Inc. for funding. F.S. and T.B. thank Toyota Research Institute for funding. T.L., M.B., S.T., and T.B. thank the U.S. Department of Energy's Office of Energy Efficiency and Renewable Energy (EERE) under the Solar Energy Technology Office (SETO) Award Number DE-EE0010503.

**Contributions**

T.L. and K.J. conceived the work. K.J. developed the methodology. T.L. collected and curated the *A*-site cation dataset. K.J. and T.L. designed the LLM prompting strategy. T.L. and F.S. performed the optimization experiment and synthesized candidate materials. F.S. collected the XRD data. T.L., F.S., and K.J. analyzed the XRD results. K.J. designed and performed the association analysis. S.T. fabricated and characterized the solar cell devices. K.J. visualized the results. K.J. and T.B. supervised the project. T.B. provided resources and funding. T.L. and K.J. wrote and all authors contributed to the edits of the manuscript.

## Competing interests

The authors declare that they have no competing interests.

## References

1. Gregory, R. L. Perceptions as hypotheses. *Philos. Trans. R. Soc. Lond. B Biol. Sci.* **290**, 181–197 (1997).
2. Laudan, L. *Science and Hypothesis: Historical Essays on Scientific Methodology*. (Springer, 2013).
3. Poincaré, H. *The Foundations of Science: Science and Hypothesis, The Value of Science, Science and Method*. (DigiCat, 2022).
4. Park, Y. J. *et al.* Can ChatGPT be used to generate scientific hypotheses? *J. Materiomics* **10**, 578–584 (2024).
5. Jing, X. *et al.* Scientific hypothesis generation process in clinical research: a secondary data analytic tool versus experience study protocol. 2022.05.21.22275060 Preprint at https://doi.org/10.1101/2022.05.21.22275060 (2022).
6. Raccuglia, P. *et al.* Machine-learning-assisted materials discovery using failed experiments. *Nature* **533**, 73–76 (2016).
7. Popper, K. R. *Conjectures and Refutations: The Growth of Scientific Knowledge.* (Psychology Press, 2002).
8. Kell, D. B. & Oliver, S. G. Here is the evidence, now what is the hypothesis? The complementary roles of inductive and hypothesis-driven science in the post-genomic era. *BioEssays* **26**, 99–105 (2004).
9. Ziatdinov, M. A. *et al.* Hypothesis Learning in Automated Experiment: Application to Combinatorial Materials Libraries. *Adv. Mater.* **34**, 2201345 (2022).

10. Ghosh, A., Kalinin, S. V. & Ziatdinov, M. A. Discovery of structure–property relations for molecules via hypothesis-driven active learning over the chemical space. *APL Mach. Learn.* **1**, 046102 (2023).

11. Liu, Y. *et al.* Autonomous scanning probe microscopy with hypothesis learning: Exploring the physics of domain switching in ferroelectric materials. *Patterns* **4**, (2023).

12. Liu, Y., Zhao, T., Ju, W. & Shi, S. Materials discovery and design using machine learning. *J. Materiomics* **3**, 159–177 (2017).

13. Wei, J. *et al.* Machine learning in materials science. *InfoMat* **1**, 338–358 (2019).

14. George, J. & Hautier, G. Chemist versus Machine: Traditional Knowledge versus Machine Learning Techniques. *Trends Chem.* **3**, 86–95 (2021).

15. Angello, N. H. *et al.* Closed-loop transfer enables artificial intelligence to yield chemical knowledge. *Nature* **633**, 351–358 (2024).

16. Butler, K. T., Davies, D. W., Cartwright, H., Isayev, O. & Walsh, A. Machine learning for molecular and materials science. *Nature* **559**, 547–555 (2018).

17. Pollice, R. *et al.* Data-Driven Strategies for Accelerated Materials Design. *Acc. Chem. Res.* **54**, 849–860 (2021).

18. Häse, F., Roch, L. M. & Aspuru-Guzik, A. Next-Generation Experimentation with Self-Driving Laboratories. *Trends Chem.* **1**, 282–291 (2019).

19. Flores-Leonar, M. M. *et al.* Materials Acceleration Platforms: On the way to autonomous experimentation. *Curr. Opin. Green Sustain. Chem.* **25**, 100370 (2020).

20. Wu, J. *et al.* Inverse design workflow discovers hole-transport materials tailored for perovskite solar cells. *Science* **386**, 1256–1264 (2024).

21. Oviedo, F., Ferres, J. L., Buonassisi, T. & Butler, K. T. Interpretable and Explainable Machine Learning for Materials Science and Chemistry. *Acc. Mater. Res.* **3**, 597–607 (2022).

22. Zhong, X. *et al.* Explainable machine learning in materials science. *Npj Comput. Mater.* **8**, 1–19 (2022).

23. Roscher, R., Bohn, B., Duarte, M. F. & Garcke, J. Explainable Machine Learning for Scientific Insights and Discoveries. *IEEE Access* **8**, 42200–42216 (2020).

24. Gunning, D. *et al.* XAI—Explainable artificial intelligence. *Sci. Robot.* **4**, eaay7120 (2019).
25. Moriconi, R., Deisenroth, M. P. & Sesh Kumar, K. S. High-dimensional Bayesian optimization using low-dimensional feature spaces. *Mach. Learn.* **109**, 1925–1943 (2020).
26. Diwale, S. *et al.* Bayesian optimization for material discovery processes with noise. *Mol. Syst. Des. Eng.* **7**, 622–636 (2022).
27. Rajabi-Kochi, M., Mahboubi, N., Singh Gill, A. P. & Mohamad Moosavi, S. Adaptive representation of molecules and materials in Bayesian optimization. *Chem. Sci.* **16**, 5464–5474 (2025).
28. Peng, P. *et al.* Bayesian optimization and explainable machine learning for High-dimensional multi-objective optimization of biodegradable magnesium alloys. *J. Mater. Sci. Technol.* **238**, 132–145 (2025).
29. German, J. *et al.* Association between plausible genetic factors and weight loss from GLP1-RA and bariatric surgery. *Nat. Med.* 1–8 (2025) doi:10.1038/s41591-025-03645-3.
30. Cipriani, V. *et al.* Rare disease gene association discovery in the 100,000 Genomes Project. *Nature* 1–9 (2025) doi:10.1038/s41586-025-08623-w.
31. Kang, K. *et al.* Study design features increase replicability in brain-wide association studies. *Nature* **636**, 719–727 (2024).
32. Tian, Y. *et al.* High-entropy hybrid perovskites with disordered organic moieties for perovskite solar cells. *Nat. Photonics* **18**, 960–966 (2024).
33. Folgueras, M. C., Jiang, Y., Jin, J. & Yang, P. High-entropy halide perovskite single crystals stabilized by mild chemistry. *Nature* **621**, 282–288 (2023).
34. OpenAI *et al.* GPT-4o System Card. Preprint at https://doi.org/10.48550/arXiv.2410.21276 (2024).
35. Maaten, L. van der & Hinton, G. Visualizing Data using t-SNE. *J. Mach. Learn. Res.* **9**, 2579–2605 (2008).
36. Song, Y. *et al.* High-Entropy Design for 2D Halide Perovskite. *J. Am. Chem. Soc.* **146**, 19748–19755 (2024).
37. Wei, J. *et al.* Chain-of-Thought Prompting Elicits Reasoning in Large Language Models. Preprint at https://doi.org/10.48550/arXiv.2201.11903 (2023).
38. Steyvers, M. *et al.* What large language models know and what people think they know. *Nat. Mach. Intell.* **7**, 221–231 (2025).

39. Peng, D. *et al.* Large-language models facilitate discovery of the molecular signatures regulating sleep and activity. *Nat. Commun.* **15**, 3685 (2024).
40. Landis, J. R. & Koch, G. G. The Measurement of Observer Agreement for Categorical Data. *Biometrics* **33**, 159–174 (1977).
41. Uffelmann, E. *et al.* Genome-wide association studies. *Nat. Rev. Methods Primer* **1**, 59 (2021).
42. Manolio, T. A. *et al.* Finding the missing heritability of complex diseases. *Nature* **461**, 747–753 (2009).
43. Fisher, R. A. Statistical Methods for Research Workers. in *Breakthroughs in Statistics: Methodology and Distribution* (eds. Kotz, S. & Johnson, N. L.) 66–70 (Springer, New York, NY, 1992). doi:10.1007/978-1-4612-4380-9_6.
44. Liu, B. *et al.* Polydentate Ligand Reinforced Chelating to Stabilize Buried Interface toward High-Performance Perovskite Solar Cells. *Angew. Chem. Int. Ed.* **63**, e202317185 (2024).
45. Tan, S. *et al.* Spontaneous formation of robust two-dimensional perovskite phases. *Science* **388**, 639–645 (2025).
46. Pang, Z., Zhang, J., Cao, W., Kong, X. & Peng, X. Partitioning surface ligands on nanocrystals for maximal solubility. *Nat. Commun.* **10**, 2454 (2019).